\documentclass[twocolumn,showpacs,aps,prl,superscriptaddress,floatfix,preprintnumbers]{revtex4}

\usepackage{graphicx}
\usepackage{dcolumn}
\usepackage{epsfig}
\usepackage{subfigure}
\usepackage{psfrag}

\input babarsym

\RequirePackage{xspace}
\usepackage{relsize}

\def\myprl  #1 #2 #3 {\jprl{#1},\ #2 (#3)}
\def\myplb  #1 #2 #3 {\plb{#1},\ #2 (#3)}
\def\myprd  #1 #2 #3 {\jprd{#1},\ #2 (#3)}
\def\mynim  #1 #2 #3 {\nim{#1},\ #2 (#3)}
\def\mypr   #1 #2 #3 {\pr{#1},\ #2 (#3)}

\newcommand{\rhop}{\ensuremath{\rho^+}\xspace}
\newcommand{\rhom}{\ensuremath{\rho^-}\xspace}
\newcommand{\rhoz}{\ensuremath{\rho^0}\xspace}
\newcommand{\rhopm}{\ensuremath{\rho^\pm}\xspace}

\newcommand{\aonez}{\ensuremath{a_1^0}\xspace}
\newcommand{\aonepm}{\ensuremath{a_1^\pm}\xspace}
\newcommand{\aeff}{\ensuremath{\alpha_\mathrm{eff}}\xspace}
\newcommand{\dd}{\ensuremath{\mathrm{d}}\xspace}

\begin{document}

\preprint{BABAR-PUB-06/052}
\preprint{SLAC-PUB-11981}

\title{ \boldmath Measurements of branching fraction, polarization, and
  charge asymmetry of $\Bpm{\to}\rhopm\rhoz$ and a search for
  $\Bpm{\to}\rhopm f_0(980)$ }

\date{\today} 

\begin{abstract}
  We measure the branching fraction ($\BR$), polarization ($f_L$) and
  \CP asymmetry ($A_{CP}$) of $\Bpm{\to}\rhopm\rhoz$ decays and search
  for the decay $\Bpm{\to}\rhopm f_0(980)$ based on a data sample of
  231.8 million \Y4S $\to$ \BB\ decays collected with the \babar\
  detector at the SLAC PEP-II asymmetric-energy $B$ factory.  In
  $\Bpm{\to}\rhopm\rhoz$ decays we measure $\BR=(16.8\pm 2.2\pm
  2.3)\times10^{-6}$, $f_L=0.905\pm 0.042 ^{+0.023}_{-0.027}$, and
  $A_{CP}=-0.12\pm 0.13\pm 0.10$, and find an upper limit on the
  branching fraction of $\Bpm{\to}\rhopm f_0(980)(\to\pip\pim)$ decays
  of $1.9\times 10^{-6}$ at 90\% confidence level.
\end{abstract}


%
\author{B.~Aubert}
\author{M.~Bona}
\author{D.~Boutigny}
\author{F.~Couderc}
\author{Y.~Karyotakis}
\author{J.~P.~Lees}
\author{V.~Poireau}
\author{V.~Tisserand}
\author{A.~Zghiche}
\affiliation{Laboratoire de Physique des Particules, IN2P3/CNRS et Universit\'e de Savoie,
 F-74941 Annecy-Le-Vieux, France }
\author{E.~Grauges}
\affiliation{Universitat de Barcelona, Facultat de Fisica, Departament ECM, E-08028 Barcelona, Spain }
% ...
\collaboration{The \babar\ Collaboration}
\noaffiliation

\maketitle


The measurement of the \CP-violating phase of the
Cabibbo-Kobayashi-Maskawa (CKM) quark-mixing matrix~\cite{CKM} is an
important part of the present program in particle physics. Violation
of \CP symmetry is manifested as a non-zero area of the CKM unitarity
triangle~\cite{Jarlskog}.  In this paper we report the measurement of
the branching fraction, polarization and \CP\ asymmetry of the
$\Bpm{\to}\rhopm\rhoz$ decay mode, which is needed for the $\rho\rho$
isospin analysis used to extract $\alpha =
\arg[-V_{td}V_{tb}^*/V_{ud}V_{ub}^*]$ \cite{gronau90}. We also set an
upper limit on the unknown branching fraction of $\Bpm{\to}\rho^\pm
f_0(980)(\to\pip\pim)$, which is measured to control this background
to the $B^\pm{\to}\rho^\pm\rhoz$ analysis.

In $\Bz(\Bzb){\to}\rhop\rhom$ decays \cite{chargeConj} the interference
between the \BB\ oscillations which depend on $V_{td}$ and the
dominating tree-level amplitude $\b\to\u\ubar\d$ causes a time-dependent
\CP asymmetry that depends on $\sin(2\alpha)$. The presence of loop
(penguin) amplitudes leads to a shift $\delta\alpha=|\alpha -
\aeff^{\rho\rho}|$, between the physical weak phase $\alpha$, and the
effective one $\aeff^{\rho\rho}$, experimentally measured in
$\Bz{\to}\rhop\rhom$ decays \cite{rhoprhom_babar,rhoprhom_belle}.
However, the penguin amplitudes in these decays are known to contribute
at a very low level due to the small upper limit of $1.1 \times 10^{-6}$
at 90\% confidence level (CL) \cite{bad1054}, obtained from the
branching fraction of the penguin dominated mode $\Bz{\to}\rho^0\rho^0$.
The size of $\delta\alpha$ can be extracted from the full isospin
analysis combining all $\B{\to}\rho\rho$ modes \cite{gronau90}.

In $B{\to}\rho\rho$ decays, a spin zero particle decays into two spin one
particles.  The final state is therefore a superposition of two
transversely polarized modes (helicity $\pm1$) and one longitudinal mode
(helicity 0), which can be measured through an angular analysis. The
longitudinal polarization fraction $f_L$ is defined as the fraction of
decays to the helicity zero state, $f_L=\Gamma_L/\Gamma$, where $\Gamma$
is the total decay rate and $\Gamma_L$ is the decay rate to the
longitudinally polarized final state.  The transverse polarization is a
mixed \CP\ state while the longitudinal state is pure \CP\ even. The
previous measurements of $f_L$~\cite{babar:rhoprhoz,belle:rhoprhoz}
showed the decay is consistent with being fully longitudinally
polarized.

Our analysis is performed in the helicity frame \cite{kramer} as a
function of the two helicity angles $\theta_\pm$ and $\theta_0$ where
the helicity angle of a \rhopm (\rhoz) meson is defined as the angle
between its daughter \pipm (\pip) 
and the direction opposite to the \B meson in the $\rhopm$ $(\rhoz)$
rest frame.  The polarization $f_L$ can be extracted from the
differential decay rate:
\begin{eqnarray}
{1 \over \Gamma} { {\dd^2 \Gamma}\over {\dd\cos\theta_\pm\
\dd\cos\theta_0} } & = & \frac{9}{4} 
\left[ f_L \cos^2\theta_\pm\cos^2\theta_0 \vphantom{1\over1}\right.\\
& & \left. + \frac{1}{4}(1-f_L)\sin^2\theta_\pm\sin^2\theta_0\right].
\nonumber
\end{eqnarray}
Here we integrate over the angle between the $\rho$-meson decay planes.

The measurements presented in this paper are based on data collected
with the \babar\ detector~\cite{detector} at the SLAC PEP-II
asymmetric-energy \epem collider. The analyzed data sample of
$231.8\pm2.6$ million \BB\ pairs produced at the $\Y4S$ resonance
corresponds to an integrated luminosity of $210.5\invfb$.


To reconstruct $\Bpm{\to}\rhopm\rhoz$ and $\Bpm{\to}\rhopm f_0$ decays,
we select events with at least three charged tracks and one neutral pion
candidate. Charged tracks are required to originate from the interaction
point and have particle identification information inconsistent with
kaon, electron, and proton hypotheses.  We form $\piz{\to}\gamma\gamma$
candidates from pairs of calorimeter showers, each with a photon-like
lateral spread and a minimum energy of 50\mev.  The invariant mass of
\piz\ candidates is required to fall in the range
$0.10<\m_{\gamma\gamma}<0.16\gevcc$.

The mass of charged $\rhopm$ candidates must satisfy
$0.396<\m_{\pipm\piz}<1.146\gevcc$ where the low-side requirement on the
$\pipm\piz$ mass is chosen to exclude $\KS{\to}\pip\pim$ decays. Neutral
final state meson candidates ($\rhoz$,\,$f_0$) must satisfy
$0.520<\m_{\pip\pim}<1.146\gevcc$.  In order to suppress backgrounds
with low momentum pions, the helicity angles are required to fall in the
ranges $-0.8<\cos\theta_\pm<0.95$ and $|\cos\theta_0|<0.95$. Backgrounds
from $\Dz\to\Km\pip\piz$ and $\Dz\to\pim\pip\piz$ decays are reduced by
requiring the candidate \Dz invariant mass to be at least $40\mevcc$
away from the \Dz\ mass.

About 20\% of the selected events have multiple \B\ candidates and the
one that has the reconstructed \piz\ mass closest to the \piz\
mass is kept.  In the case that more than one candidate has the same
reconstructed \piz\ mass, we select one at random.

Continuum decays represent the largest source of background and are
reduced by requiring $|\cos\theta_T|<0.8$, where $\theta_T$ is the
cosine of the angle between the \B thrust axis and that from the rest of
the event (ROE).  To further discriminate signal from continuum, we also
use a neural network built out of five event-shape variables: a Fisher
discriminant combining the $0^{th}$ and $2^{nd}$ order monomials
\cite{pipiBabar} for charged particles and neutral clusters of the ROE;
the cosine of the angle between the direction of the \B and the
collision axis ($z$) in the center-of-mass (CM) frame; the cosine of the
angle between the \B thrust axis and the $z$ axis; the variable
$|\cos\theta_T|$ defined above; and the sum of transverse momenta in the ROE
relative to the $z$ axis. The output is transformed into a variable
$x_{NN}$ which has roughly Gaussian signal and background distributions.
We select candidates in a range of $x_{NN}$ that removes 54\,\% of
continuum background events while retaining 90\,\% of the signal. After
these selections, about 85\% of the remaining events are from continuum
decays.

Signal event candidates are further identified based on two kinematic
variables: the beam-energy-substituted mass $\mes=\sqrt{(s/2+{\bf
    p}_i\cdot{\bf p}_B)^2/E_i^2-{\bf p}_B^2}$, using the total initial
\epem\ 4-momentum $(E_i,{\bf p}_i)$, CM energy~$(\sqrt{s})$ and the
\B\ momentum~$({\bf p}_B)$, and $\DeltaE = E^{*}_{B}-\sqrt{s}/2$, the
difference between the reconstructed \B\ energy in the CM
frame~$(E^{*}_{B})$ and the beam energy. Events are selected if
$\mes>5.26\gevcc$ and $|\DeltaE|<150\mev$.

After the selection criteria are applied, the efficiency is $8.4\,\%$
for longitudinal and $18.6\,\%$ for transverse polarized
$\Bpm{\to}\rhopm\rhoz$ decays. The selection efficiency is $16.6\,\%$
for $\Bpm{\to}\rhopm f_0$ decays.
Any possible interference effects between the $\Bpm{\to}\rhopm\rhoz$ and 
$\Bpm{\to}\rhopm f_0$ are neglected.


An unbinned extended maximum likelihood fit is applied to the selected
sample of $N_\mathrm{tot}=74293$ events in order to measure the $\Bpm{\to}\rhopm\rhoz$
event yield, polarization, and charge asymmetry as well as the
$\Bpm{\to}\rhopm f_0$ event yield. The likelihood function is:
\begin{equation}
{\cal L} = \frac{1}{N_{\rm tot}!}\exp\left(-\sum_{k=1}^{M} n_{k}\right)\, 
\prod_{i=1}^{N_{\rm tot}} 
\left(\sum_{j=1}^{M}~n_{j}\, 
{\cal P}_{j}(\vec{x}_{i})\right),
\label{eq:likel}
\end{equation}
where $M$ is the number of hypotheses (signal, mis-reconstructed signal,
continuum and \B-background classes), and $n_k$ ($n_j$) represents the
number of measured events for each hypothesis determined by maximizing
the likelihood function.  ${\cal P}_{j}(\vec{x}_{i})$ is the product of
the probability density functions (PDFs) of hypothesis $j$ evaluated at
the $i$-th event's measured variables, $\vec{x_i}=\{\mes$, $\DeltaE$,
$m_{\pipm\piz}$, $m_{\pip\pim}$, $\cos\theta_\pm$, $\cos\theta_0$,
$x_{NN}\}$.  In addition, the charge asymmetry, obtained from the
measured \Bm\ and \Bp\ signal candidate decay yields,
$A_{CP}=\frac{N_{B^-}-N_{B^+}}{N_{B^-}+N_{B^+}}\ ,$ is determined in the
fit to the data.

Each discriminating variable in the likelihood function is modeled with a PDF
extracted either from the data, or from high statistics Monte Carlo (MC) simulated data samples.
The correlations between the variables are assumed to be small and the
PDFs independent. This is checked with systematic error studies, and
corrections are applied where necessary. 

The continuum background \DeltaE, \mes, and $x_{NN}$ distributions are
modeled with one-dimensional parameterized distributions taken from
fits to the data.  Correlations are observed between the $m_{\pi\pi}$
and $\cos\theta$ distributions for both $\rho$-meson candidates, which
are taken into account with two-dimensional PDFs. The signal component
is modeled with one-dimensional parameterized distributions for each of
six variables; \mes is modeled with a non-parametric PDF~\cite{keys}.
The signal PDF shapes are obtained from fits to signal MC sample after
the selection is applied.  Events with a true $\Bpm{\to}\rhopm\rhoz$
decay but with wrong tracks or calorimeter clusters assigned to the
final state are referred to as self cross feed (SCF) events. They make
up 35\% and 14\% of the selected longitudinally and transversely
polarized signal samples, respectively.  The longitudinal and transverse
SCF components and \B-background PDFs are determined in a similar manner
using high statistics MC samples and modeled with non-parametric
PDFs~\cite{keys} for each variable.


To understand the backgrounds from other \B\ decay modes we use MC
simulated events. There are two types of \B-background: `charmed'
(decays involving $b{\to}c$ transitions), and `charmless' (all other $b$
decays). Altogether sixteen \B-background categories plus the two SCF
components are included in the fit. The SCF yields and polarization are
fixed in the final fit at values that match those fitted for the signal
in previous iterations of the fit.  Four specific charmed background
modes are included: $\Bm{\to}\Dz\pim$, $\Bm{\to}\Dz\rhom$,
$\Bm{\to}\Dstarz\pim$, and $\Bm{\to} \Dstarz\rhom$.  Other charmed
backgrounds are combined into two generic classes of events for charged
and neutral charmed \B\ decays. For the charmless \B-backgrounds,
separate MC samples of eight modes were used: neutral $B$ decaying to
$\rhop\rhom$ and charged $B$ decaying to $\rhopm f_0(980)$,
$\eta'\rhopm$, $\Kstarz\rhopm$, $\aonez\pipm$, $\aonepm\piz$,
$a_1^\pm\rho^0$, and $a_1^0\rhopm$ with the decays $a_1^0\to(\rho\pi)^0$
and $a_1^\pm\to(\rho\pi)^\pm$.  For $B$ decaying to vector-mesons, only
the longitudinal component of the decay is considered.  Two generic
categories, one for 5-body modes and one for all `other charmless'
decays, complete the \B-background model.

The number of `other charmless' events and the $\Bpm\to\rhopm f_0$ yield
were determined from the data fit. The other fourteen backgrounds had
their yields fixed in the fit. We use the following branching fractions:
\begin{tabular}{rlc}
$\BR(\Bz\to\rhop\rhom)$&$=(26.2\pm3.7)\times 10^{-6}$\ \  &\cite{hfag_lp_2005}, \\
$\BR(\Bpm\to\eta'\rhopm)$&$=(12.9\pm6.5)\times 10^{-6}$ &\cite{babar:BtoetaX}, \\
$\BR(\eta'\to\rhoz\gamma)$&$=0.295\pm0.010$ &\cite{pdg04},\\
\hspace{0.8cm}$\BR(\Bp\to\Kstarz\rhop)$&$=(10.5\pm1.8)\times 10^{-6}$ &\cite{hfag_lp_2005}, \\
\multicolumn{3}{c}{and $\BR(\Kstarz\to\Kp\pim)=2/3$.} \\
\end{tabular}
\newline
The decays $\Bpm{\to}(a_1\pi)^\pm$ and $\Bpm{\to}(a_1\rho)^\pm$ have few
experimental constraints \cite{babar:a1pi,babar:conf_a1rho}. We adopt the
following $\Bpm$ branching ratios, in units of $10^{-6}$, 
and assume a 100\% systematic
uncertainty: $a_1^0\pip = 12$, $a_1^+\piz = 6$, $a_1^0\rhop  =
a_1^+\rho^0  = 48$.

\begin{figure}[!ht]
\begin{center}
 \resizebox{0.48\textwidth}{!}{
{\label{fig:mes}\includegraphics[width=0.35\textwidth]{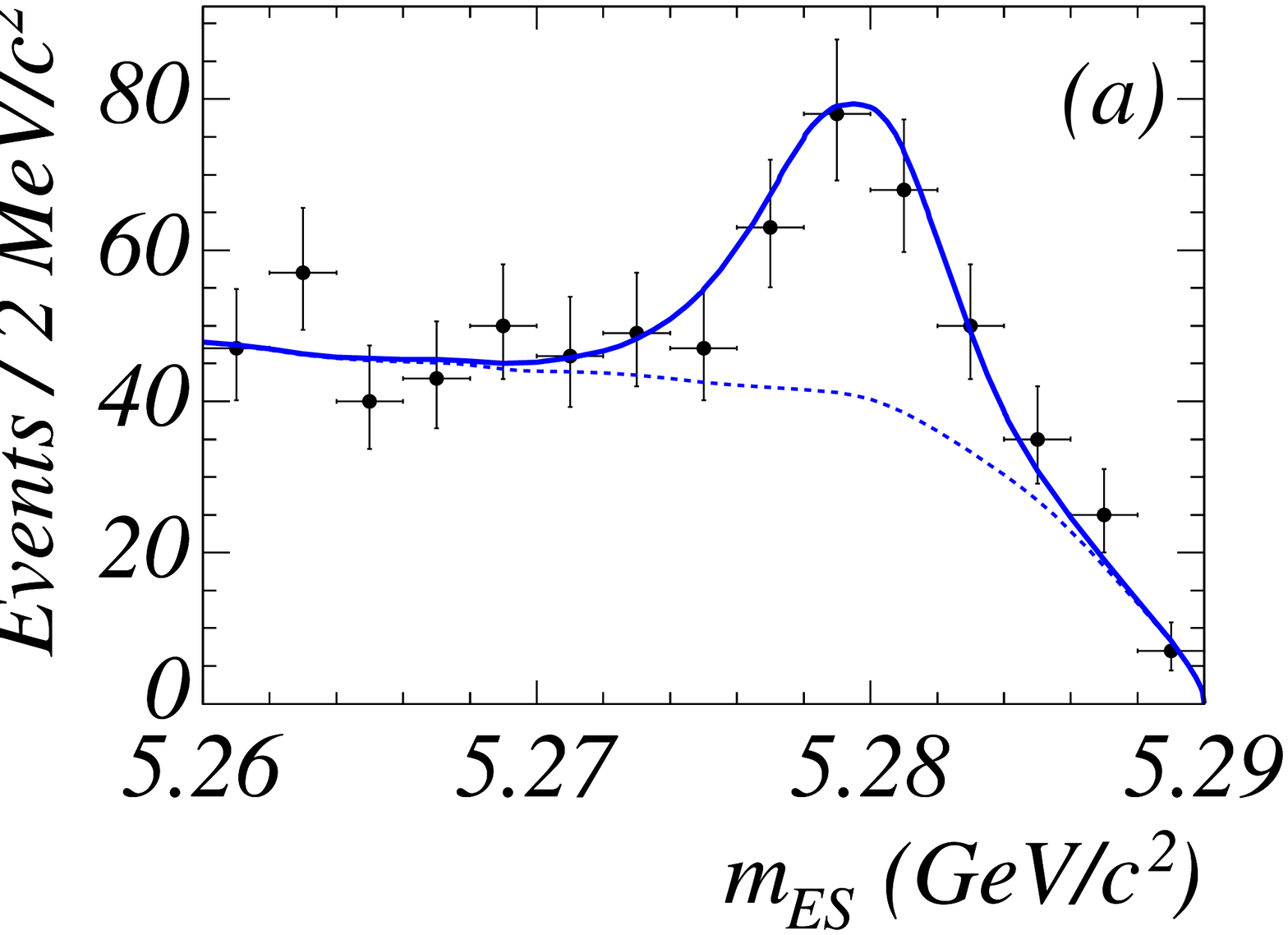}}
\hspace{0.01\textwidth}
{\label{fig:de}\includegraphics[width=0.35\textwidth]{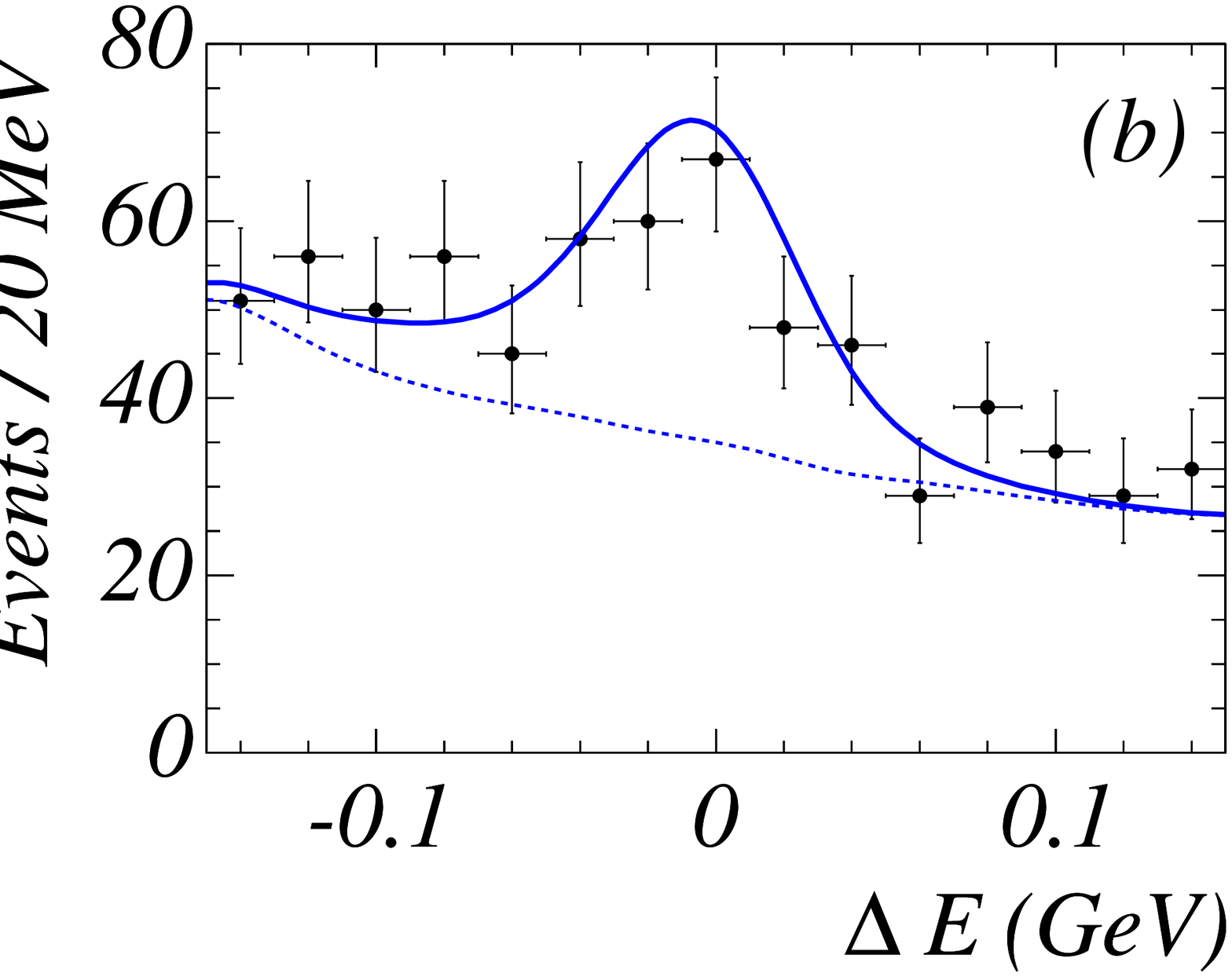}}
}
\vspace{-0.2in}
 \resizebox{0.48\textwidth}{!}{
   {\label{fig:mrhop}\includegraphics[width=0.35\textwidth]{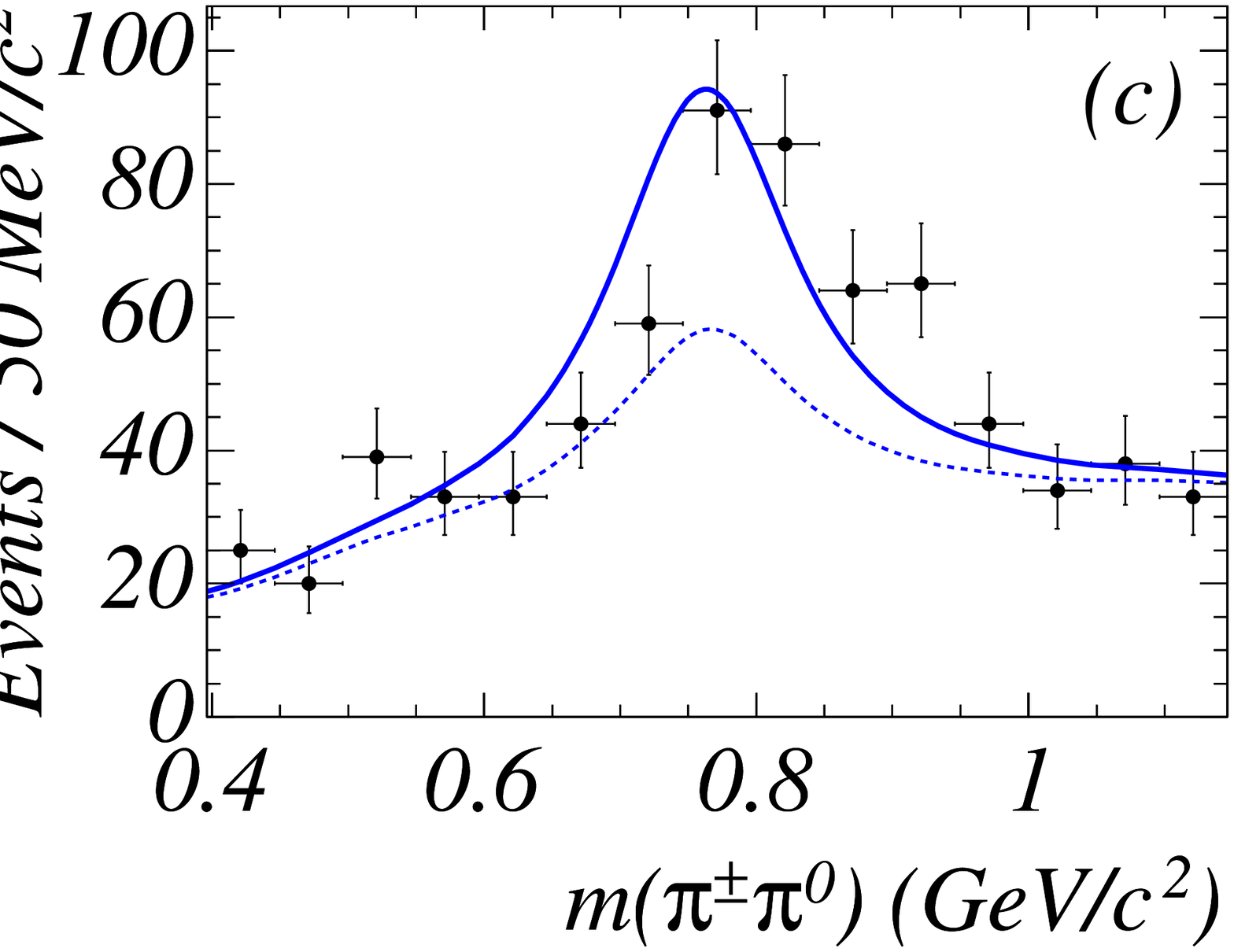}}
 \hspace{0.01\textwidth}
  {\label{fig:mrhoz}\includegraphics[width=0.35\textwidth]{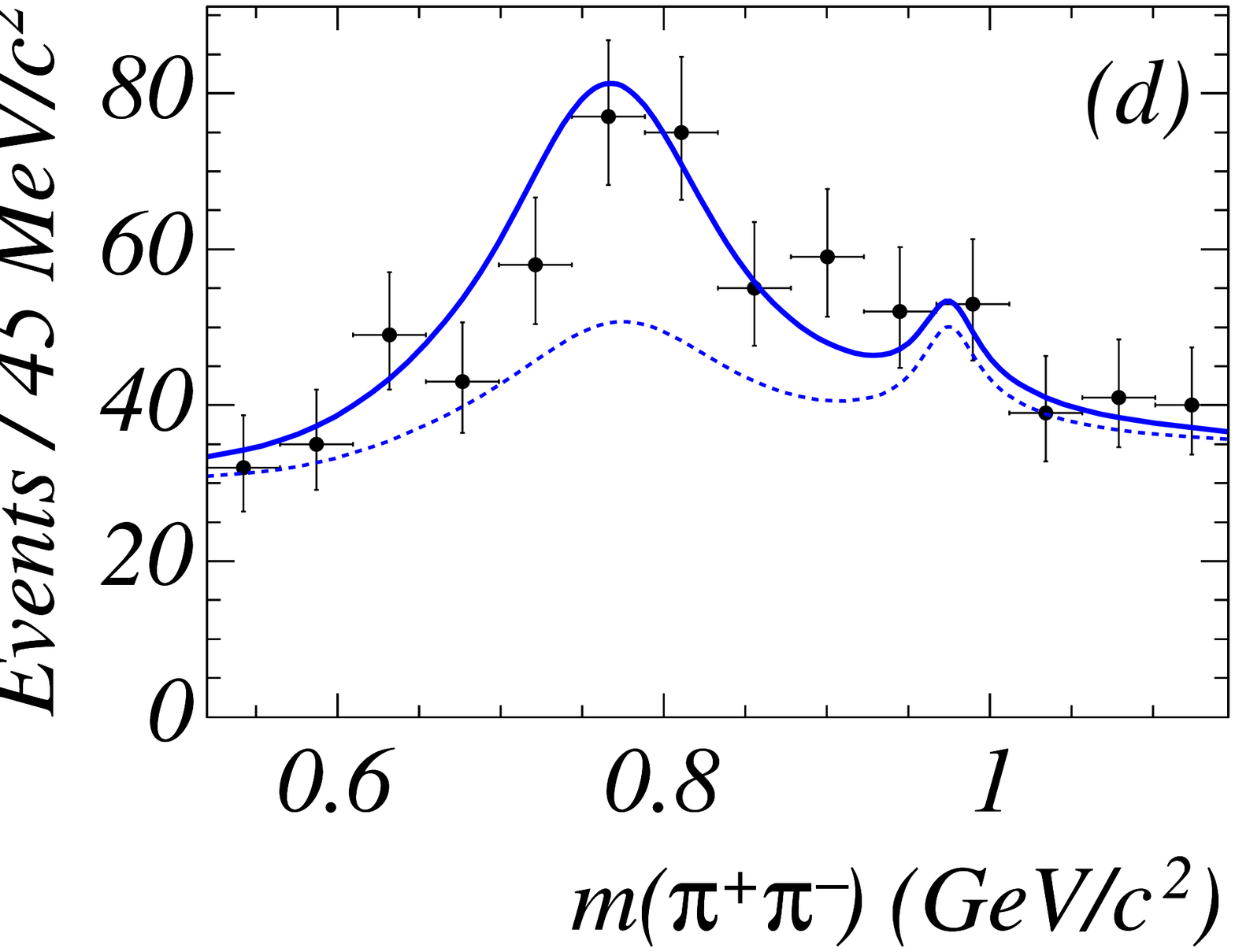}}
 }
\end{center}
\caption{Projections of (a) $\mes$, (b) $\Delta E$, (c)
$m_{\pipm\piz}$, and (d) $m_{\pip\pim}$ with a cut on the ratio of the
signal and background likelihoods that selects about 40\% of the
signal. For (a) and (b), the observable plotted is excluded from the
fit in calculating the likelihood used for the enrichment
selection. For (c) and (d), only the $\mes$, $\Delta E$, and $x_{NN}$
variables have been used in calculating the likelihood.  Points
represent on-resonance data, dashed lines the continuum and
$B\overline{B}$ backgrounds PDFs, and solid lines the likelihood
function with yields taken from the fit were all variables have been
used.}
\label{fig:projplots}
\end{figure}

Table~\ref{table:results} shows the results of the fit, where the quoted errors
are statistical errors only. Projection plots for $\mes$ and $\DeltaE$ are 
shown in Fig.~\ref{fig:projplots}.

\begin{table}[!ht]
  \caption{Summary of the results of the fit with statistical errors
  (before correction for fit biases).\label{table:results}}
  \begin{center}
    \begin{tabular}{cc}      
      \hline
      \hline
      \multicolumn{1}{c}{Observables} & Fitted value \\
      \hline
      $\Bpm\to\rhopm\rhoz$ yield & 390$\pm$49 events\\
      Polarization $f_L$ & 0.897$\pm$0.042 \\
      Charge asymmetry $A_{CP}$ &  $-0.12\pm0.13$ \\
      $\Bpm\to\rhopm f_0$ yield & 51$\pm$30 events\\
      \hline
      \hline
    \end{tabular}
    \vspace{-0.15cm}
  \end{center}
\end{table}

Systematic effects are considered in the modeling of the PDF shapes and
biases in the fit model. Tests of the fit are made by using a large number
of MC samples containing the amounts of signal, continuum and \B-background
events measured or fixed in the data fit.
The fits to these samples should reproduce the number of MC events
modeled. A shift of the fitted values with respect to the generated
ones indicates a bias in the fitting procedure, for which a correction
to the yield is then applied.  We use the same technique to study the
effects of correlations between the neural net and helicity variables in
the $q\overline{q}$ continuum.
No other significant correlations were observed between the other discriminating variables.
After
combining these two effects we find an intrinsic bias in the fit of $+25\pm
27$ on the number of $\Bpm{\to}\rhopm\rhoz$ events, $-0.008\pm 0.005$ on
$f_L$ and $+23.6\pm9.8$ events on the $\Bpm{\to}\rhopm f_0$ yield.  These
biases are corrected from the fit measurements, and half of each
separate fit bias is taken as the systematic error.

Many of the \B-background rates are poorly known. The effect of
uncertainties in these values is evaluated by varying the number of
events in each background category within the range allowed by the error
on the branching fraction.  Fourteen non-resonant backgrounds that are
not in the default fit are tested by adding them singly to the fit with
a yield that is allowed to vary. The only shift seen was associated with
the mode $B^\pm\rightarrow \pi^\pm\pi^0\pi^0$, and is taken as a
symmetric systematic uncertainty.

The systematic error associated with mis-reconstructed signal is
evaluated by taking the difference between the default fit and the one
for which these events are not modeled, and therefore mostly absorbed
into the `other charmless' background category.  We consider the error
due to the uncertainty on the signal, \B-background, and continuum PDF
shapes and estimate a systematic error by varying these shapes within
their statistical uncertainty. The impact of the uncertainty on the
measurement of the $f_0$ mass and width~\cite{f0} has also been
evaluated. The values of the systematic errors described above are given
in Table~\ref{table:syst}.

\begin{table}[!ht]
  \caption{Summary of the systematic uncertainties on the
    $\Bpm{\to}\rhopm\rhoz$ yield, 
    the polarization $f_L$, and the $\Bpm{\to}\rhopm f_0$ yield.
    \label{table:syst}}
  \begin{center}

    \begin{tabular}{lccc}
      \hline
      \hline
 Source    & $\rhopm\rhoz$ yield & $f_L$ & $\rhopm f_0$ yield \\
      \hline
Fit bias      & 27.3& 0.005 & 9.8\\
\B-background rates & 11.0 & 0.007 & 2.8\\
Non-resonant backgrounds &12.0 & 0.009 & 3.0\\
Amount of SCF & 24.0 & 0.010 & 0.6\\
PDF shapes & $^{+21.1}_{-22.5}$ & $^{+0.017}_{-0.022}$ & $^{+7.9}_{-13.5}$\\
$f_0$ mass and width &$^{+0.9}_{-0.6}$ &0.000 & 3.9\\
 \hline
Total & $^{+45}_{-46}$ &$^{+0.023}_{-0.027}$ & $^{+14}_{-18}$ \\
      \hline
      \hline
    \end{tabular}
  \end{center}
\end{table}

Systematic uncertainties in the reconstruction and calibration procedure
introduce a systematic error of $3\%$ after a correction of $-2.5\%$ on
the \piz\ reconstruction efficiency, $3.9\%$ after a correction of
$-1.5\%$ on the track reconstruction efficiency, and a systematic error
of $1.1\%$ from the particle identification. The uncertainty on the
efficiency ratio between longitudinal and transverse events is found to
be negligible. The error on $A_{CP}$ includes a $0.45\%$ uncertainty in
the charged track reconstruction asymmetry, a $4\%$ uncertainty from the
detector's intrinsic charged particle identification asymmetry, and a
$9\%$ uncertainty which is the largest single shift obtained when
assuming a uniform probability for the charge asymmetry of every
\B-background individually.


In summary, we measure the branching fraction, longitudinal
polarization, and \CP asymmetry of the decay
$\Bpm\rightarrow\rhopm\rhoz$, using a dataset of about 231.8 million
$B\overline{B}$ pairs, to be:
\begin{eqnarray}
\BR(\Bpm\to\rhopm\rhoz)&=&(16.8\pm 2.2\pm2.3)\times10^{-6}, \nonumber \\ 
f_L(\Bpm\to\rhopm\rhoz)&=&0.905\pm0.042^{+0.023}_{-0.027}, \nonumber \\ 
A_{CP}(\Bpm\to\rhopm\rhoz)&=&-0.12\pm0.13\pm 0.10. \nonumber
\end{eqnarray}
The measurement of the branching fraction has improved by a factor of
about two with respect to the previous \babar\ measurement
\cite{babar:rhoprhoz}, and supersedes it. The isospin relations
between branching ratios are consistent between this measurement and
those of $\rhop\rhom$ and $\rhoz\rhoz$ \cite{hfag_lp_2005}, validating
the approach \cite{gronau90} used to constraint $\alpha$.  Moreover,
our measurements confirm that this mode is largely longitudinally
polarized. They also confirm that the charge asymmetry is consistent
with zero as expected for decays proceeding through one decay channel
only; this suggests the contributions of electroweak penguins are
small in the $\B{\to}\rho\rho$ system.

In addition we measure $\BR(\Bpm\to\rhopm
f_0(980)(\to\pip\pim))=(0.7\pm0.8\pm0.5)\times10^{-6}$ with a
significance of $0.4\sigma$. We set an upper limit on the branching
fraction of $1.9\times10^{-6}$ at 90\% confidence level by finding the
yield ($N$) that satisfies $\int_0^N{\cal L}(n)dn/\int_0^\infty{\cal
  L}(n)dn=0.9$ taking into account systematic uncertainties.

We are grateful for the excellent luminosity and machine conditions
provided by our \pep2\ colleagues, 
and for the substantial dedicated effort from
the computing organizations that support \babar.
The collaborating institutions wish to thank 
SLAC for its support and kind hospitality. 
This work is supported by
DOE
and NSF (USA),
NSERC (Canada),
IHEP (China),
CEA and
CNRS-IN2P3
(France),
BMBF and DFG
(Germany),
INFN (Italy),
FOM (The Netherlands),
NFR (Norway),
MIST (Russia), and
PPARC (United Kingdom). 
Individuals have received support from the
Marie Curie EIF (European Union) and
the A.~P.~Sloan Foundation.



\begin{thebibliography}{99}
\bibitem{chargeConj} Charge conjugation is implied throughout this
  document, unless explicitly stated.
\end{thebibliography}
\end{document}